\documentclass[twocolumn,showpacs,preprintnumbers,amsmath,amssymb,
superscriptaddress]{revtex4}

\usepackage{graphicx}

\usepackage{epsfig}

\usepackage{dcolumn}

\usepackage{bm}

\usepackage{natbib}

\begin{document}

\title{Accurate Model of a Vertical Pillar Quantum Dot}

\author{P.A. Maksym}

\affiliation{Department of Physics and Astronomy, University of Leicester,
Leicester LE1 7RH, UK}

\author{Y. Nishi}
\affiliation{Department of Physics, University of Tokyo, Hongo, Bunkyo-ku,
Tokyo 113-0033, Japan}

\author{D.G. Austing}

\affiliation{Institute for Microstructural Sciences M50, National Research
Council, Montreal Road, Ottawa, Ontario K1A 0R6, Canada}

\author{T. Hatano}

\affiliation{ICORP Spin Information Project, JST, Atsugi-shi,
Kanagawa 243-0198, Japan}

\author{L.P. Kouwenhoven}

\affiliation{Department of Applied Physics, Delft University of Technology,
P.O.Box 5046, 2600 GA Delft, The Netherlands}

\author{H. Aoki}

\affiliation{Department of Physics, University of Tokyo, Hongo, Bunkyo-ku,
Tokyo 113-0033, Japan}

\author{S. Tarucha}

\affiliation{ICORP Spin Information Project, JST, Atsugi-shi,
Kanagawa 243-0198, Japan}

\affiliation{Department of Applied Physics, University of Tokyo, Hongo,
Bunkyo-ku, Tokyo 113-8656, Japan}

\date{\today}


\begin{abstract}
An accurate model of a vertical pillar quantum dot is described.
The full three dimensional structure of the device containing the dot is
taken into account and this leads to an effective two dimensional model
in which electrons move in the two lateral dimensions, the confinement is
parabolic and the interaction potential is very different from the bare
Coulomb potential. The potentials are found from the device structure and a
few adjustable parameters. Numerically stable calculation procedures for
the interaction potential are detailed and procedures for deriving
parameter values from experimental
addition energy and chemical potential data are described. The model is able
to explain magnetic field dependent addition energy
and chemical potential data for an individual dot to an accuracy of about
5\%, the accuracy level needed to determine ground state quantum
numbers from experimental transport data. Applications to excited state
transport data are also described.
\end{abstract}

\maketitle

\section{Introduction}
Vertical pillar quantum dots exhibit an extremely wide range of interesting
physics, particularly in the presence of a magnetic field perpendicular to
the plane of the dot \cite{Tarucha96,Kouwenhoven97,Oosterkamp99}. The
Coulomb blockade allows single electrons to be added to the dot so the
number of electrons can be set precisely, starting from one. For each
number of electrons the ground state evolves through a series of
transitions as the magnetic field is increased. At zero magnetic field
the ground states are well described by Hund's first rule and
maximum density droplet (MDD) states occur when the field is increased to a
few Tesla. Finally, fractional quantum Hall droplet and electron molecule
states occur at high fields beyond about 10T. So, roughly
speaking, there are four regimes but the detailed
picture is much richer. Many transitions occur in between each of the main
regimes and each transition is accompanied by abrupt changes in the spin
and orbital angular momentum of the ground state. There is experimental
\cite{Tarucha96,Kouwenhoven97,Oosterkamp99,Kouwenhoven01,Nishi06} and
theoretical \cite{Maksym90,Maksym96,Chakraborty99,Maksym00,Nishi06}
evidence that these effects occur. However it is difficult to apply
experimental techniques, like scanning tunnelling spectroscopy, to probe the
corresponding quantum states directly. Instead the ground state quantum
numbers are found by comparing data from transport spectroscopy with 
calculated results. This requires an accurate dot model that can be used to
analyse data for an {\it individual} quantum dot. Development of an
appropriate model is the purpose of the present work.

Although quantum dots are often described as artificial atoms, there is a very
important difference between an artificial atom and a natural one: all
natural atoms of the same isotope are identical but all quantum dots are
different. This makes it quite challenging to model an individual dot
accurately, even when its basic design parameters are known, because 
manufacturing tolerances introduce fluctuations in the parameters and
this can have a significant influence on dot states. The fact that the
dot often consists of a small region embedded in a much larger device
structure just adds to the difficulty of constructing an accurate model.
The ideal model of an electrostatic quantum dot 
is a system in which electrons are constrained to
move in the two dimensions parallel to the dot plane, are confined by a
parabolic potential and interact via the Coulomb interaction. The
properties of this model have been examined extensively, see
\cite{Chakraborty99} for a review. It gives a good qualitative description
of dot behaviour but a more accurate model is needed for data analysis.
One possibility is a device model in which the dot confining
potential and in some cases the interaction potential are computed,
together with the quantum states, from a combined solution of the Poisson
and Schr\"odinger equations. This approach has been used by many authors,
\cite{Bruce00,Matagne02,Bednarek03,Melnikov05} for example, and has given 
a great deal
of insight into the generic behaviour of devices containing quantum dots.
However, it is extremely difficult to use generic device
models to deduce ground state quantum numbers from experimental data for an
{\it individual} quantum dot.

One of the difficulties encountered in analysing individual dot data
is that energies have to be computed to very high precision.
The quantity measured experimentally is the gate voltage at which transport
occurs. This is
proportional to the chemical potential, $\mu_N$ which is the 
difference of two dot energies: $\mu_N = E_N - E_{N-1}$, where $E_N$ is the
energy of the an $N$-electron dot state. Further, the
transport data is often presented as a difference between the gate
voltages for two successive current peaks. This gives the addition energy,
$E_{AN} = E_{N+1} - 2 E_N + E_{N-1}$, the second difference of $E_N$ with
respect to $N$. Because energy differences are needed to compare
theoretical results with
experimental data the energies themselves must be computed very
accurately. Typically, the addition energy is around 3 meV while typical
dot energies are one or two orders of magnitude larger so that high precision
values of the dot energy are needed for data analysis. However the potentials
found in generic dot models depend on many parameters, dot dimensions,
dopant densities etc, some of which are not known accurately. In principle,
these parameters could be adjusted to fit the data but it is desirable to
avoid fitting a large number of parameters.

The present model lies between the parabolic confinement, Coulomb
interaction model and the generic device models. The main idea is to
develop a model that contains all the essential physics but depends on a
small number of adjustable parameters. Three are used in the present work
but it turns out that one of them is not very significant. The model is a
parabolic confinement model, believed to be accurate for small numbers of
electrons \cite{Matagne02}. Two of the parameters determine the
parabolic potential but only one of them is significant. The interaction 
potential is determined from a
simplified device structure. The effects of finite thickness and
screening on the interaction in this structure are included fully and
a third model
parameter is used to determine the screening. Essentially, the resulting model
reduces to a two-dimensional, parabolic confinement model with an interaction
potential that is very different from the bare Coulomb potential. The model
gives an excellent description of addition energy data, accurate to about
0.15 meV. It has already been used to determine the ground state spin and
orbital angular momentum of a pillar dot in the strong magnetic field
regime \cite{Nishi06}. However ref.~\cite{Nishi06} does not contain the
detailed explanation of the model and analysis procedures that is given
here.

The model is described in section \ref{ModelSection} and the method used to
calculate the quantum states is detailed in section \ref{CalcSection}.
Following this, parameter fitting procedures for experimental addition
energy and chemical potential data are discussed in section
\ref{FitSection}. Then
the application of the model to data in the low and high magnetic field
regimes is described in section \ref{ResultsSection} and conclusions are
stated in section \ref{ConcSection}. Finally, two appendices deal with the
calculation of Green's functions and interaction matrix elements needed to
find the quantum states.

\section{Dot Model}
\label{ModelSection}
\subsection{Vertical pillar dots}
\begin{figure}
\begin{center}
\includegraphics[width=7.5cm,angle=0]{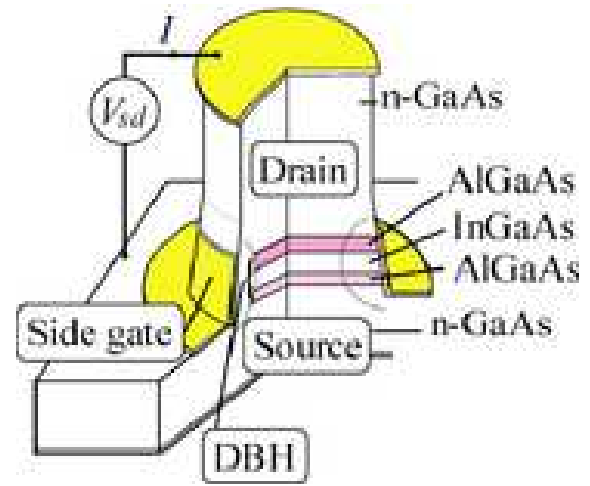}
\includegraphics[width=7.5cm,angle=0]{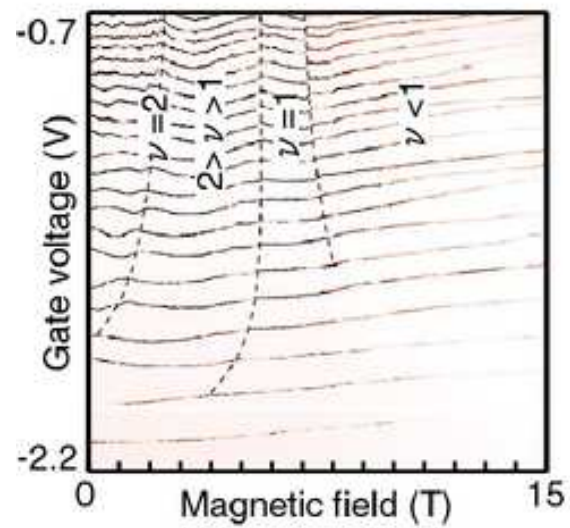}
\end{center}
\caption{Typical vertical pillar dot (top) and typical transport data
(bottom). Labels indicate the Landau level filling factor $\nu$.}
\label{dotfig}
\end{figure}
A vertical pillar dot (Fig. \ref{dotfig}, top frame) consists of a
narrow pillar, typically 500 nm in diameter \cite{Kouwenhoven01}. The
pillar contains a double barrier heterostructure (DBH) which
provides confinement in the vertical ($z$) direction and is surrounded by a
cylindrical metal gate which carries a negative charge that provides
lateral ($x,y$) confinement. The pillar stands on a substrate which is
heavily doped and there is also a heavily doped region at the top of the
pillar. These heavily doped regions provide source and drain contacts that
enable the transport properties of the device to be investigated. A unique
advantage of the vertical geometry is that it is not sensitive to edge
states which affect transport through lateral devices in the high magnetic
field regime.

In a typical experiment, the source-drain current is measured as a function
of the gate voltage, $V_g$, source-drain voltage, $V_{sd}$ and magnetic field,
$B$ parallel to the pillar. Typical results \cite{Nishi06} are shown in the
bottom frame of Fig. \ref{dotfig}. This image shows the condition for an
electron to enter the dot at fixed source drain bias. It is well known that
transport through the device is Coulomb blockaded except at certain values
of $V_g$ and $V_{sd}$
\cite{Kouwenhoven01}. In Fig.~\ref{dotfig} the current is colour coded so
that dark regions correspond to the largest current. The $N$\/th dark curve 
shows
how the gate voltage needed for the $N$\/th electron to enter the dot
depends on the magnetic field. The lowest curve in the figure
corresponds to $N=1$.

The exact condition for electron transport is
\begin{equation}
\mu_c + \frac{eV_{sd}}{2} \ge \mu_N - e \alpha(V_{gN}) V_{gN} \ge
\mu_c - \frac{eV_{sd}}{2},
\end{equation}
where $\mu_c$ is the contact chemical potential, $V_{gN}$ is the gate
voltage at which the $N$\/th electron enters the dot
and $\alpha$ is an electrostatic leverage factor.
At zero source drain bias this reduces to
\begin{equation}
e \alpha(V_{gN}) V_{gN} = \mu_N - \mu_c
\label{mudef}
\end{equation}
so that the gate voltage at which
transport occurs is a measure of the dot chemical potential relative to the
contact. For transport data in the form of a gate voltage difference,
$V_{gN} - V_{gN-1}$ is proportional to the addition energy,
$E_{AN} = E_{N+1} - 2 E_N + E_{N-1}$ if $\alpha$ and $\mu_c$ are independent of
$N$. Hence the first and second differences of the dot energies are needed
to compare theory with experiment.

\subsection{Overview of model}
The dot model takes account of vertical confinement, lateral confinement
and interactions between dot electrons. The quantum well between the double
barriers is relatively narrow (12 nm) so the electrons are confined to the
ground state of the vertical motion. This leads to a quasi-two-dimensional
model in which the interaction is modified by the vertical confinement,
physically the effect is to smear out the Coulomb singularity.

The lateral
confinement is generated by the cylindrical side gate and in principle the
lateral confinement potential may be found by solving the Poisson equation,
refs. \cite{Bruce00,Matagne02,Bednarek03,Melnikov05}, for example. However 
in the present case, $N$ is small so the spatial extent
of the dot state ($<50$ nm) is very small compared to the pillar diameter
($\approx 500$ nm). Therefore the only part of the confining potential that
has a significant effect on the dot electrons is the part near the
minimum which is parabolic. In addition, observation of shell structure in
the device considered here \cite{Nishi06} shows that the device has a high
degree of circular symmetry. The lateral confining potential is therefore
taken to
be $V_{c} = V_0 + m^*\omega_0^2 r^2/2$, where $r$ is the lateral radial
co-ordinate and $\omega_0$ is found from $\hbar \omega_0 = a + b(N-1)$ where
$a$ and $b$ are fitting parameters. The constant $V_0$ is not needed for 
the present analysis because the addition energy is independent of $V_0$
and the chemical potential is also independent of $V_0$ when $\mu_N$ taken
relative to $\mu_1$. Previous work has shown that
the parabolic approximation is consistent with experiment \cite{Maksym90}
and accurate for small numbers of electrons \cite{Matagne02}, that is
electrons in the first and second shells at $B=0$.

The screening is mainly caused by heavily doped contacts that are
relatively close
to the dot, around 10 nm from the well. This is close enough for the
contact regions to have a significant screening effect on the interaction
between the dot electrons. This is treated in the Thomas-Fermi
approximation and the screening length in the contact region is taken to be
a fitting parameter. Full details of the present approach to 
screening and finite thickness are given in the following two sections.

\subsection{Finite thickness}
\label{FTSection}
The dot is three dimensional but when the electrons are confined to the
vertical ground state their motion is quasi-two-dimensional. A
variational method is used to find the resulting effective Hamiltonian.
In principle this needs two steps. The envelope function for the
vertical confinement should be found first and then the spin splitting
should be found from this envelope function \cite{Lommer85}. However the
second step is not needed here because the experimental value of 
the effective $g$-factor for the present device is known. Except for the
spin splitting terms, the 3D effective mass Hamiltonian is
\begin{eqnarray}
H &=& \sum_i p_{zi} \frac{1}{2 m^*(z)} p_{zi} +
{\bm \pi}_i \frac{1}{2 m^*(z)} {\bm \pi}_i + V_c(r_i) + V_b \Theta(z_i)
\nonumber\\
&+& \frac{1}{2}\sum_{i\ne j} U({\bf r}_i-{\bf r}_j, z_i, z_j),
\end{eqnarray}
where $\mbox{\boldmath{$\pi$}} = {\bf p}_\parallel + e{\bf A}$,
$\bf{p}_\parallel$ is the lateral momentum, ${\bf A}$ is the magnetic vector
potential, $p_z$ is the vertical momentum,
$V_b$ is the barrier height, $\Theta(z)$ is a step function,
$\Theta(z)=1$ when $|z|>w/2$, $\Theta(z)=0$ otherwise, $w$ is the well
width, $m^*(z) = m^*_w$ when $z$ is in the well and $m^*(z) = m^*_b$ when
$z$ is in the barrier. $U$ is the electron-electron interaction potential
and ${\bf r} = (x,y)$. The variational trial function for the electron 
states is taken to have the
form $\Psi = \Pi_i \chi(z_i)\Phi({\bf r}_1,... {\bf r}_N)$, where $\Phi$
is antisymmetric and includes spin functions. 

Equations
for $\Phi$ and $\chi$ are found by minimising $\langle\Psi|H|\Psi\rangle$
subject to the constraint that $\Phi$ and $\chi$ are normalised. This
leads to
\begin{eqnarray}
\bar{H}_\parallel |\Phi\rangle = \Lambda |\Phi\rangle, \nonumber \\
\left[ H_\perp + \frac{E_\parallel(m^*_b)-E_\parallel(m^*_w)}{N} \Theta
\right]|\chi\rangle = \Lambda' |\chi\rangle \label{eigeneq2},
\end{eqnarray}
where $\Lambda$ and $\Lambda'$ are eigenvalues,
$m^*_w$ and $m^*_b$ are effective masses in the well and barrier
respectively and $\bar{H}_\parallel = p_w H_\parallel(m^*_w) +
(1-p_w) H_\parallel(m^*_b)$. Here $p_w$ is the probability of finding and
electron in the well, $p_w = \int^{w/2}_{-w/2} \chi^2(z)dz$ and the
parallel and perpendicular Hamiltonians are defined by
\begin{eqnarray}
H_\parallel(m) &=& \sum_i \frac{\pi^2_i}{2 m} + V_c(r_i) \nonumber \\
&+& \frac{1}{2}\sum_{i\ne j}
\int \chi^2(z) \chi^2(z') U({\bf r}_i-{\bf r}_j, z, z') dz dz',
\nonumber \\
H_\perp &=& p_z \frac{1}{2 m^*(z)} p_z + V_b \Theta(z).
\label{Hdefs}
\end{eqnarray}
The parallel energies in the second of Eqs.~(\ref{eigeneq2}) are
expectation values of $H_\parallel$: $E_\parallel(m) =
\langle\Phi|H_\parallel(m)|\Phi\rangle$. After Eqs.~(\ref{eigeneq2}) have
been solved the total energy is found from the expectation value of the
full Hamiltonian:
\begin{equation}
E= N \langle\chi|H_\perp|\chi\rangle +
\langle\Phi|\bar{H}_\parallel|\Phi\rangle.
\label{Etot}
\end{equation}

In principle, the coupled eigenvalue problem defined by
Eqs.~(\ref{eigeneq2}) can be solved iteratively to arbitrary accuracy
but in practice this is not necessary. The energy difference term
$[E_\parallel(m^*_b)-E_\parallel(m^*_w)]/N$ in the second of
Eqs.~(\ref{eigeneq2}) in the present case is around 1\% of the barrier
height. Hence the effect of this term is small and it is sufficient to find
$\chi$ from the zeroth-order approximation in which the energy difference
term is neglected and then use the zeroth-order $\chi$ to find the averaged
Hamiltonian $\bar{H}_\parallel$. The averaging has two physical effects.
First, it changes the effective mass. The dot considered here is made from
InGaAs containing 5\% In, and the In reduces the mass. The barrier is made
from AlGaAs containing 22\% Al and this increases the mass. The net effect is
to reduce the mass to $0.0653 m_0$, slightly less than the GaAs value. The
second effect of the averaging is to smear out the Coulomb singularity in
the effective two-dimensional interaction. This results from the integral
over $z$ in the first of Eqs.~(\ref{Hdefs}).
The averaging procedure also applies to systems with
spin-orbit coupling. In this case, it is found that the Dresselhaus
spin-orbit coupling
parameter has to be averaged in a way similar to the effective mass. This
has been used for studies of spin relaxation in the present device
\cite{Chaney07}.

The final step in determining the effective Hamiltonian is to find the
spin splitting. The effective $g$-factor, $|g^*|$, is determined from 
the experimentally observed Zeeman splitting \cite{Nishi06} in the same 
device that is used for the data analysis performed here.  $|g^*| = 0.3$
for the dot when $B \agt 10$ T, compared with $|g^*| =
0.44$ for bulk GaAs. The reduced value for the dot is consistent with the 
effect of non-parabolicity \cite{Lommer85}. The non-parabolicity also
makes the effective $g$-factor depend on magnetic field and Landau level
index \cite{Lommer85} but the resulting corrections to $\mu_N$ are
probably smaller than experimental error. A constant $g$-factor is
therefore used and the Zeeman energy is accounted for by adding
$g^*\mu_B B S_z$ to the total energy in Eq.~\ref{Etot}. Here $\mu_B$ is
the Bohr magneton and
$S_z$ is the $z$ component of the total spin $S$. The effective
$g$-factor is assumed to be negative, as in bulk GaAs, so $g^*=-0.3$.

\subsection{Screening}

The most important source of screening in the present device is the
heavily doped contacts above and below the dot. The dielectric response of
the disordered contact material is not well understood but can be treated
approximately in the static Thomas-Fermi approximation. This allows the
screening length to be estimated from the density of states at the Fermi
level, however this is not known accurately for the contact material. Hence
the screening length is taken to be a fitting parameter which is determined
from experimental data.

The screened interaction between dot electrons \cite{Hallam96} is found
from the
electrostatic Green's function $G({\bf R},{\bf R}')$ which satisfies
\begin{eqnarray}
-\mbox{\boldmath{$\nabla$}} \cdot \left[
(\epsilon({\bf R})\epsilon_0 \mbox{\boldmath{$\nabla$}}
G({\bf R},{\bf R}')\right] &=& -q_0^2 ({\bf R}) \epsilon({\bf R}) \epsilon_0
G({\bf R},{\bf R}') \nonumber \\
&+& \delta({\bf R}-{\bf R}'),
\label{3DGreen}
\end{eqnarray}
where $\epsilon({\bf R})$ is the relative permittivity,
$q_0 = 2 \pi / \lambda_s$ inside the contacts, $q_0 =0$ outside the
contacts and ${\bf R} = ({\bf r},z)$. Because a detailed theory of magnetic
field dependent screening by the 3D contact material is not available, the
screening length, $\lambda_s$, is taken to be independent of the magnetic 
field. This approximation should be good when the screening length does not
change significantly over the field range used in the experiments and it is
assumed that this is the case. Eq.~(\ref{3DGreen})
is simplified by making use of the fact that the dot is an
order of magnitude smaller than the pillar. This means that the Green's
function is almost translationally invariant in the lateral direction and
can be found to a good approximation by replacing the pillar by a stack of
dielectric layers of infinite extent in the lateral directions. $\epsilon$
and $q_0$ then do  not depend on ${\bf r}$. So the Green's function
is translationally invariant in the lateral directions and can be expressed
as a Fourier transform:
\begin{equation}
G({\bf R},{\bf R}') = \frac{1}{4\pi^2} \int G({q},z,z')
\exp(i{\bf q}\cdot({\bf r}-{\bf r}')) d{\bf q}.
\end{equation}
The Fourier components of the Green's function satisfy
\begin{eqnarray}
&-&\frac{d}{dz}\left[ \epsilon(z)\epsilon_0 \frac{d}{dz}
G({q},z,z')\right] \nonumber \\ &+&
\epsilon(z)\epsilon_0 [q^2 + q^2_0(z)]G({q},z,z') = \delta(z-z').
\label{gfode}
\end{eqnarray}
The solution of this equation has the form
\begin{eqnarray}
G({q},z,z') = -\frac{f(z)g(z')}{W}, \qquad z > z',\nonumber \\
G({q},z,z') = -\frac{g(z)f(z')}{W}, \qquad z < z',
\label{gfform}
\end{eqnarray}
where $f$ and $g$ are solutions of the homogeneous equation corresponding to
Eq.~(\ref{gfode}) and $W$ is their Wronskian. These solutions are chosen to
satisfy $f \rightarrow 0$ when $z \rightarrow +\infty$
and $g \rightarrow 0$ when $z \rightarrow -\infty$. Although the analytic
form of $G$ is as given in Eq.~(\ref{gfform}), severe numerical
instabilities are encountered when this form is evaluated unless special
precautions are taken. The problem is that the $G$ has rapidly growing
components which cause exponent overflow. Exactly the same mathematical
problem is
encountered in calculations of evanescent wave propagation in electron
diffraction theory and the solution used here is a reflection matrix
approach developed for calculations of reflection high energy electron
diffraction (RHEED) \cite{Meyer-Ehmsen89,Zhao93,Kawamura07}. This is
detailed in appendix A.

\begin{figure}
\begin{center}
\includegraphics[width=5.5cm,angle=-90]{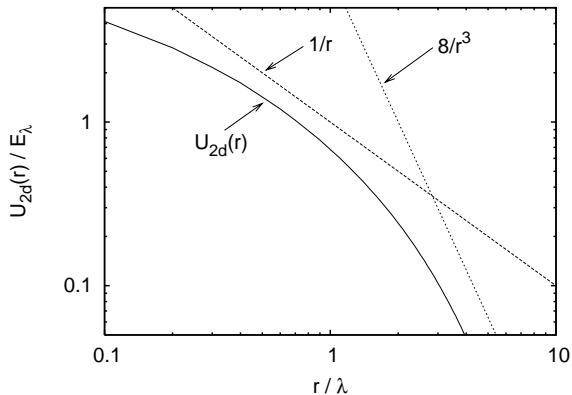}
\end{center}
\caption{Effective 2D interaction (solid line). Dashed lines show the
$1/r^3$ limiting form and the $1/r$ Coulomb interaction.}
\label{ufig}
\end{figure}

Once the Green's function has been found, the Fourier components of the
effective two-dimensional interaction are obtained in the form of an
integral:
\begin{equation}
U_{2d} (q) = e^2 \int \chi^2(z) \chi^2(z') G({q}, z, z') dz dz'.
\label{u2dqdef}
\end{equation}
The special form of the Green's function, Eq.~(\ref{gfform}), is used to
simplify this integral so that it can be evaluated efficiently.
Details are given appendix B. The form of the effective interaction in real
space is very different from a pure Coulomb interaction. Because of the
screening, the interaction is dipole-like at long range and decreases
like $1/r^3$ in the limit of large $r$. And the Coulomb singularity is
removed by the finite thickness because the electrons are able to move out
of the dot plane and avoid each other. The real-space effective interaction
as a function of $r$ is shown in Fig.~\ref{ufig}. $U_{2d}(r)$ is calculated
numerically from its Fourier transform $U_{2d}(q)$ for the layer structure
shown in Fig.~\ref{dotfig} (see also Table~\ref{laytab}) and
$\lambda_s = 10$ nm,
a typical screening length. The length unit is the 2D harmonic oscillator
length parameter, $\lambda^2 = \hbar / (2 m^* \Omega)$ where
$\Omega^2 = \omega_0^2 + \omega^2_c/4$ with $\omega_c$ the cyclotron
frequency. The energy unit $E_{\lambda} = e^2 /
(4\pi\epsilon_w\epsilon_0\lambda)$ where $\epsilon_w$ is the relative
permittivity in the well that contains the dot.
It is clear that $U(r)$ is significantly different from the $1/r$
Coulomb interaction and approaches
the $1/r^3$ form in the limit of large $r$.

The effect of the dielectric interfaces in the device on the effective
interaction is accounted for fully in the Green's function formalism. In
addition, the dielectric interfaces have an effect on the vertical
confinement potential. This results from the interaction of each electron
with its own image charges and is also accounted for in the Green's
function formalism. The image charge contribution to the vertical
confinement is obtained from the non-singular part of the Green's function
\cite{Hallam96} and in the present case the image charge contribution is
\begin{eqnarray}
V_I(z) &=& \frac{e^2}{8\pi\epsilon_w\epsilon_0}
\int^{\infty}_0 \left[2q\epsilon_w\epsilon_0G(q,z,z) - 
\frac{\epsilon_w}{\tilde{\epsilon}} \right] d{q},
\end{eqnarray}
where $\tilde{\epsilon} = \epsilon_w$ when $z$ is in the well, 
$\tilde{\epsilon} = \epsilon_b$, the barrier relative permittivity, 
when $z$ is in the barrier and
$\tilde{\epsilon} = (\epsilon_w+\epsilon_b)/2$ when $z$ is at the interface
between the well and the barrier. The image charge term leads  
to a small change in the vertical confinement which is taken
into account via perturbation theory.

Another source of screening in the present device is the metallic,
cylindrical side gate. In principle, this breaks the translational
invariance of the effective lateral interaction however this effect is
expected to be small. The magnitude of the effect can be estimated from
the electrostatic Green's function for an infinite metallic cylinder
\cite{Smythe68}. Numerical calculations of the Green's function suggest that
the effect is $\alt 1$\% within a region of radius $\sim 10$ nm in the
centre of the cylinder. The effect is clearly largest for electrons at the
edge of the dot but the interaction between these electrons and electrons
in the centre of the dot is suppressed by the screening effect of the
contacts. These considerations suggest that the effect of the gate is small
for the small electron numbers considered here ($N\le 5$). However,
the effect could be significant for larger electron
droplets whose edge is closer to the metallic gate.

\subsection{Model Hamiltonian}
\begin{table}
\caption{Layer structure of the device model.}
\begin{ruledtabular}
\begin{tabular}{lcr}
Layer  & Composition & Thickness\\
\hline
Top contact & n$^+$ GaAs&\\
Buffer &i GaAs& 3 nm\\
Barrier &Al$_{0.22}$Ga$_{0.78}$As& 9 nm\\
Well &In$_{0.05}$Ga$_{0.95}$As& 12 nm\\
Barrier &Al$_{0.22}$Ga$_{0.78}$As& 7.5 nm\\
Buffer &i GaAs& 3 nm\\
Bottom contact & n$^+$ GaAs&\\ 
\end{tabular}
\end{ruledtabular}
\label{laytab}
\end{table}
\begin{table}
\caption{Material parameters for the device model.}
\begin{ruledtabular}
\begin{tabular}{lcr}
Parameter  & Material & Value\\
\hline
$m^*/m_0$ & Al$_{x}$Ga$_{1-x}$As& $0.067 + 0.083 x$\\
$m^*/m_0$ & In$_{y}$Ga$_{1-y}$As& $0.067 - 0.045 y$\\
$V_b$ (meV)& Al$_{x}$Ga$_{1-x}$As / In$_{y}$Ga$_{1-y}$As&
$823x + 640y$\\
$\epsilon_b$ &Al$_{x}$Ga$_{1-x}$As& $12.7 - 3.12x$\\
$\epsilon_w$ &In$_{y}$Ga$_{1-y}$As& $12.7 +2y$\\
\end{tabular}
\end{ruledtabular}
\label{mattab}
\end{table}
The effective Hamiltonian for the two-dimensional dot model is
\begin{eqnarray}
H_{eff} &=& \sum_i \frac{\pi^2_i}{2 \bar{m}^*} + V_c(r_i)
+ N E_z + g^*\mu_B B S_z
\nonumber \\
&+& \frac{1}{2}\sum_{i\ne j} U_{2d}({\bf r}_i-{\bf r}_j),
\end{eqnarray}
where $E_z = \langle\chi|H_\perp|\chi\rangle + \Delta V_I$ is the 
perpendicular energy, $\Delta V_I$ is the perturbation caused by the image
charge term. The averaged effective
mass $\bar{m}^*$ is found from $1/\bar{m}^* = p_w/m^*_w + (1-p_w)/m^*_b$
and the effective $g$-factor is determined experimentally (Section
\ref{FTSection}). The layer structure of the device is detailed in
Table~\ref{laytab} and the material parameters used for the present
calculations are detailed in Table~\ref{mattab}.

\section{Calculation of Eigenstates}
\label{CalcSection}
The effective Hamiltonian $H_{eff}$ is very similar to the Hamiltonian for a
two dimensional parabolic dot with a Coulomb interaction. Hence the
eigenstates of $H_{eff}$ are found in the usual way by numerical
diagonalization in a Fock-Darwin basis \cite{Chakraborty99}. The only
difference between the usual calculation and the present one is that matrix
elements of the effective interaction $U_{2d}$ have to be computed. In the
case of the Coulomb interaction these matrix elements are normally found
from the Fourier transform of the interaction and the calculation involves
an integral over $q$ \cite{Chakraborty99}. In the present case the Fourier
transform approach is also used. $U_{2d}(q)$ has the form $F(q)V_{2d}(q)$
where $V_{2d}(q)$ is the Fourier transform of the Coulomb interaction,
$e^2/(4\pi\epsilon_w\epsilon_0r)$, in
two dimensions and $F(q)$ is a form factor that is computed numerically from
the Green's function. The only difference between the standard treatment of
the interaction in a 2D dot and the present treatment is the appearance of 
the form factor in the $q$ integral. This integral is done numerically by 
Romberg integration.

The Fock-Darwin states are labelled by an angular momentum quantum number
$l$ and a radial quantum number $n$. The many-electron basis used for the
diagonalization consists of Slater determinants formed from these states.
It is important to minimise the size of this basis as expensive
calculations have to be performed repeatedly to fit the model parameters.
The Hamiltonian is block diagonalized according to the value of the total
orbital angular momentum $J$. For $N \le 4$ the basis for each $J$ is
formed from Fock-Darwin states with $n\le 3$ and all $l$ values compatible
with the required $J$ value. For $N=5$ the size of the basis is limited by
making use
of the fact that the radial excitation is an inter-Landau level excitation
and hence has large energy in a strong magnetic field. This enables the
maximum $n$ quantum number to be reduced as a function of magnetic field.
In the present case the maximum value of $n$ is taken to be the integer
part of $6.9 - 0.28 B$ which corresponds to a maximum $n$ of 6 at $B=0$ T
and 2 at $B=14$ T. All Slater determinants compatible with this $n$ value
are then constructed and those Slater determinants whose energy is within
100 meV of the lowest energy determinant are retained in the calculation.
This rejection step reduces the total number of determinants slightly and
saves some computer time. The accuracy of the numerically calculated ground
state addition energies is estimated to be 0.1 meV or better. The
$B$-dependent cut-off on $n$ leads to small steps in the energy as a
function of $B$ which result from the sudden change in the number of basis
states that occurs whenever the integer part of $6.9 - 0.28 B$ changes.
These artefacts are typically around 0.01 meV, about an order of magnitude
smaller than experimental error. 

\section{Data Analysis}
\label{FitSection}
\subsection{Parameter fitting}
\label{fitSection}
Standard techniques are used to fit the model parameters. The general idea
is to find parameters that minimise the metric distance between
experimental and theoretical curves. This problem occurs in many areas
of physics and the present work is based on a formalism developed for surface
structure analysis \cite{Stock90,Coy98}. For each curve, the metric
distance is taken to be
\begin{equation}
D = \frac{1}{N_p} \sum_i \left[ f_t(B_i) - f_e(B_i)\right]^2,
\end{equation}
where $f_t$ and $f_e$ are theoretical and observed values of the
addition energy or chemical potential at magnetic field $B_i$ and $N_p$ is
the number of experimental points in the curve. The metric distance used to
fit multiple curves is the average of the metric distances for the
individual curves weighted with the number of points in each curve.
With this choice, the procedure reduces to an unweighted least squares
fit of the entire data set. It is possible to adjust the metric to give
weight to specific features in the data, such as peaks and shoulders, but
this is not done in the present work. The metric distance is minimised with
the Levenberg-Marquardt algorithm, a standard numerical technique
\cite{Press93}. This fitting procedure is generally accepted to give
reliable parameter values provided the number of features in the data
exceeds the number of fitting parameters, a condition that is satisfied in
the present case.

\subsection{Analysis of addition energy data}
\label{addSection}
The addition energy data was collected as part of an experiment to
investigate ground state transitions in the high-magnetic field regime
\cite{Nishi06}. The data was taken with a fairly high magnetic field
resolution (0.05 T) and consists of a limited number of $V_{gN}$ curves,
$N=2,3,4,5$ only. The measurements were performed in a dilution
refrigerator and the electron temperature is estimated to be below 100
mK.

\begin{figure}
\begin{center}
\includegraphics[width=9.25cm,angle=-90]{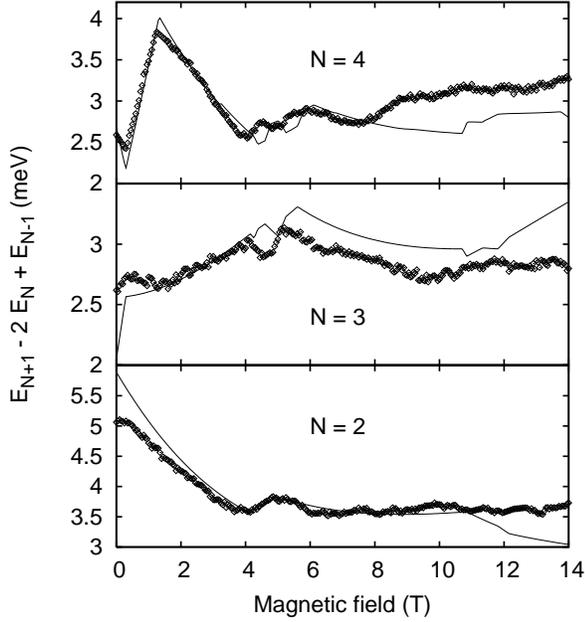}
\end{center}
\caption{Experimental and theoretical addition energies. The
field range used for parameter fitting is $0\le B \le 14$ T.
Diamonds: experimental data, solid lines: theoretical results.}
\label{addfig1}
\end{figure}

\begin{figure}
\begin{center}
\includegraphics[width=9.25cm,angle=-90]{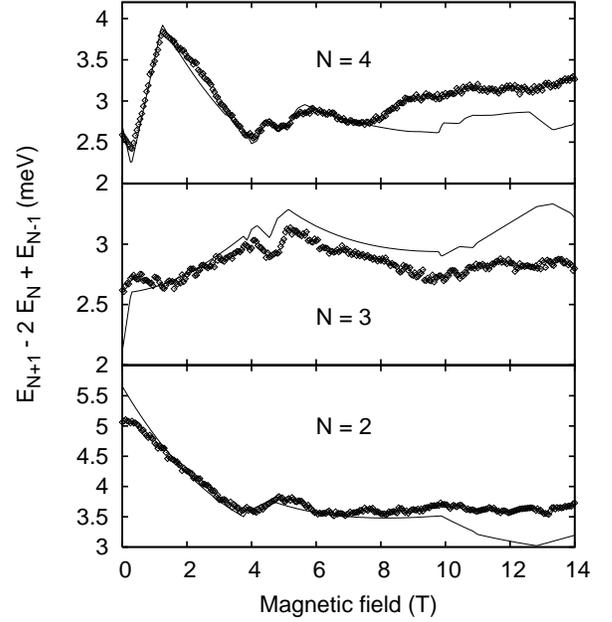}
\end{center}
\caption{Experimental and theoretical addition energies. As
Fig.~\ref{addfig1} except that the
field range used for parameter fitting is $0\le B \le 10$ T.}
\label{addfig2}
\end{figure}
The ground state addition energy is the difference of two successive
chemical potentials: $E_{AN} = E_{N+1} - 2 E_N + E_{N-1} =
\mu_{N+1} - \mu_N$. It follows from Eq.~(\ref{mudef}) that $E_{AN}$
at fixed magnetic field is related to two successive gate voltages:
\begin{equation}
E_{AN} = e[\alpha(V_{gN+1}) V_{gN+1} - \alpha(V_{gN})V_{gN}],
\label{alphaeq1}
\end{equation}
which is valid provided that the contact chemical potential $\mu_c$ is
independent of $V_g$ and therefore cancels when the $\alpha V_g$ values are
subtracted. Eq.~(\ref{alphaeq1}) can in principle be used to find
$E_{AN}$ from experimental values of $\alpha$ and $V_g$. However this
requires very accurate values of both $\alpha$ and $V_g$. Typical values of
$E_{AN}$ are around 3 meV, while typical values of $e \alpha V_g$ are
around 60 meV. So both $\alpha$ and $V_g$ need to be measured to an
accuracy of 0.1 \% or better to measure $E_{AN}$ to $\pm 0.1$ meV. 
$V_g$ is known to this accuracy for the present device but $\alpha$ is
not. Eq.~(\ref{alphaeq1}) is therefore approximated by
\begin{equation}
E_{AN} = e\bar{\alpha}_{N+1}(V_{gN+1} - V_{gN}),
\label{alphaeq2}
\end{equation}
where $\bar{\alpha}_{N+1} = [\alpha(V_{gN+1}) + \alpha(V_{gN})]/2$. This
suppresses the random errors which would result from use of
Eq.~(\ref{alphaeq1}) but is only valid when $\alpha(V_{gN})$ varies
sufficiently slowly with $N$. This is {\it assumed} to be the case. The
validity of this assumption cannot be tested without more data on $\alpha$
however the results based on this assumption are found to be consistent
with results of chemical potential analysis, section \ref{muSection}.
Another factor that
limits the present analysis is that $\alpha$ depends on the magnetic field
as well as $V_g$ but is only known for a sample of field points at
intervals of around $1-2$ T. The $\alpha$ values at intermediate field
points have to be obtained by linear interpolation and this introduces a
further systematic error which is difficult to quantify. In principle, both
this systematic error and the one resulting from use of $\bar{\alpha}$
can be eliminated by performing detailed measurements of $\alpha$.

Figure~\ref{addfig1} shows the comparison between experimental and
theoretical addition energies for $N=2$, 3, 4. All three curves were used
simultaneously to fit the parameters, as described in section
\ref{fitSection}. It is clear that the absolute values of the experimental and
theoretical data agree very well. The raw experimental data is shown in the
figure together with raw theoretical data. There is no shifting or scaling
of the theoretical curves. The agreement is generally good  but there are 
some discrepancies,
particularly around 4-6 T and above 10 T. To investigate the cause of these
discrepancies the data was fitted again in the restricted field range
$0\le B \le 10$ T. The results are shown in Fig.~\ref{addfig2} and it is
clear that the fit is better, particularly around 4-6 T where the double
peak structure for $N=3$ and $N=4$ is reproduced very well. The
improvement in the fit can be
quantified with the metric distance which is equivalent to the RMS
difference between experiment and theory.
The value corresponding to Fig.~\ref{addfig1} is 0.23 meV while for
Fig.~\ref{addfig2} it is 0.15 meV.

The deterioration in the quality of the fit when the full data range is
used could be caused by factors not included in the model or uncertainties
in the experimental data. The most likely causes are impurity effects,
screening and insufficient knowledge of $\alpha$.

Both impurity effects and screening could depend on magnetic field.
Impurity effects are not included in the present
model but are likely to be more significant in the very strong field regime.
The mean distance between the impurities in the heavily doped contacts is
around 10 nm. This gives a fluctuating contribution to the dot potential
on a similar length scale. But at zero magnetic field the typical diameter
of the dot state is around 20 - 30 nm. Hence the effect of the impurities
will tend to average out. Indeed the experimental evidence is that the
present device has a high degree of circular symmetry
\cite{Nishi06}. However as the magnetic field increases the dot wave
function shrinks and eventually becomes smaller than the length scale of
the potential fluctuations. Impurity effects would be large in this regime
but the magnetic fields used in the present work are probably too low for
this to happen. Magnetic field dependent screening is another possible
cause of discrepancies but if the magnetic field dependence was
significant, it would probably occur throughout the field range, while the
discrepancies are mainly in the high field range.

The remaining possible cause of the discrepancies is that the value of
$\alpha$ is not known to
sufficient accuracy in the high-field regime. In this case the parameters
determined by fitting the data up to 10 T are expected to provide an
accurate description of the dot in the field range up to 14 T but the
problem lies in the numerical
value of the experimental addition energy. There is some evidence that this
is the case: excited state features predicted by the present model are
found to correspond to features in experimental excitation spectra in the
high-field regime \cite{Nishi06}. The value of $\alpha$ is not needed to
identify these features and this suggests that the discrepancies
in the high field addition energy are most likely to be related to
insufficient knowledge of $\alpha$.

The best fit parameter values corresponding to Fig.~\ref{addfig2} 
are $a=4.8 \pm 0.1$ meV, $b=0.02 \pm 0.01$ meV and
$\lambda_s=15.0 \pm 3$ nm. Here the statistical errors correspond to the 
parameter
range compatible with the fluctuations in the data shown in Fig.~\ref{addfig2}
but exclude the unquantifiable uncertainty in $\alpha$.
These values are taken to give the best model of the present device. The
value of $b$ suggests that the confinement energy increases slowly with
gate voltage but the opposite would be expected from electrostatics of the
device. However the value of $b$ is comparable to its error so a model
with constant confinement or slowly decreasing confinement would
probably give an equally good fit to the data.

\subsection{Analysis of chemical potential data}
\label{muSection}
In principle, Eq.~(\ref{mudef}) can be used to obtain $\mu_N$ from the
$V_g$ data but the appearance of the contact chemical potential
$\mu_c$ in this equation presents an obstacle. The absolute value of
$\mu_c$ is not known. In addition, $\mu_c$ is expected to vary with
magnetic field and its field dependence is not known. Part of the
difficulty can be eliminated by taking the chemical potential relative to
the chemical potential for the first electron at zero magnetic field.
The contact chemical potential is written as $\mu_c(0) + \Delta \mu_c(B)$,
where $\Delta\mu_c$ vanishes when $B=0$. Eq.~(\ref{mudef}) becomes
\begin{eqnarray}
e [\alpha(V_{gN}, B) V_{gN}(B) - \alpha(V_{g1},0)V_{g1}(0)] &=&
\nonumber \\
\mu_N(B) - \mu_1(0) -\Delta \mu_c(B).
\label{mufit}
\end{eqnarray}
This enables experimental and theoretical values of $\mu_N(B) - \mu_1(0)$
to be compared
provided that a model for $\Delta \mu_c(B)$ is available. A model
investigated in the present work is $\Delta\mu_c(B) =
\sqrt{\Gamma^2 + (\hbar\omega_c/2)^2} - \Gamma$, where $\Gamma$ is a
fitting parameter. An advantage of analysing chemical potential data is
that Eq.~(\ref{mufit}) is less sensitive to errors in $\alpha$ than
Eq.~(\ref{alphaeq1}) however the disadvantage is that the results depend on
the model chosen for $\Delta \mu_c(B)$.

Chemical potential data analysis requires a data set that contains
$V_{g1}$ so that $\mu_N(B) - \mu_1(0)$ can be found. The addition energy
data analysed in section \ref{addSection} does not contain $N=1$ data so  
a second data set is used for the chemical potential data analysis. This
data was collected under
similar conditions to the addition energy data but over a wider range
of $N$, a smaller magnetic field range and a lower field resolution (0.1 T).
In addition, the device was
subjected to one thermal cycle between collecting the two data sets. The
chemical potential data was collected first then the device was taken up to
room temperature and cooled again to collect the addition energy data.

\begin{figure}
\begin{center}
\includegraphics[width=5.5cm,angle=-90]{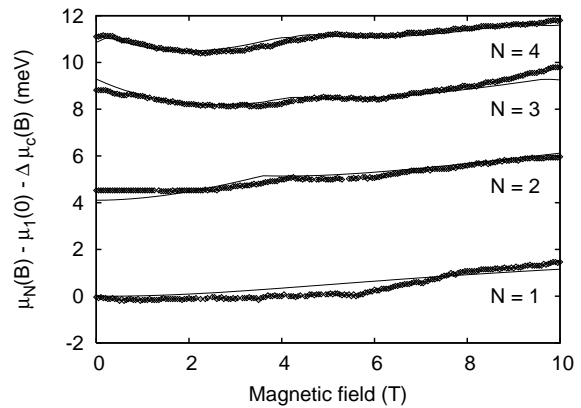}
\end{center}
\caption{Experimental and theoretical chemical potentials.
Diamonds: experimental data, solid lines: theoretical results.}
\label{mufig}
\end{figure}
The comparison between experimental and theoretical chemical potentials is
shown in Fig.~\ref{mufig}. As with the addition energy, a very good fit is
obtained without any scaling or shifting of the theoretical data. The RMS
difference between the experimental and theoretical curves is about
0.12 meV. The best fit parameter values are
$a=4.8 \pm 0.1$ meV, $b=-0.06 \pm 0.02$ meV, 
$\lambda_s=16.0 \pm 2$ nm and $\Gamma = 7.4 \pm 0.3$ meV. Except for the
sign of $b$, these values are consistent with the values obtained by
fitting the addition energy. As with the addition energy the absolute
magnitude of $b$ is small and the statistical error in $b$ is relatively
large. This suggests that the variation of $\hbar \omega_0$ with $N$ is not
very significant for the small $N$ range considered here.

The best model of the device is believed to be the model that results from
fitting the addition energy because an extra parameter is needed to fit the
chemical potential data and it is preferable to keep the number of fitting
parameters to a minimum. The addition energy model was therefore used in 
ref. \cite{Nishi06} to determine ground state quantum numbers from
transport data. However ref. \cite{Nishi06} contains excited state data 
and comparison of features in experimental and theoretical
excitation data requires comparison of chemical potential data. This
appears to be an insurmountable 
problem because the addition energy data does not contain $V_{g1}(B)$
so it is impossible to fit $\Gamma$ for the experimental conditions of
ref. \cite{Nishi06}. However, the
exact value of the chemical potential is not needed for the analysis of
ref. \cite{Nishi06} and similar situations because the only information that 
needs to be extracted from the experimental and theoretical data is the 
magnetic fields at which features occur. As an aid to
identification of these features it is very useful to compute the chemical
potential as $\mu_N(B) - \mu_1(0) - \Delta \mu_c(B)$ because this ensures
that the experimental and theoretical data have roughly the same slope.
This can be achieved by using an approximate value of $\Gamma$ that is 
found by visual comparison of the data, $8.2$ meV in the case of ref.
\cite{Nishi06}.
The precise value of $\Gamma$ is not important in this case as $\Gamma$
does not affect the position of features in the transport data.

\section{Applications of the model}
\label{ResultsSection}
\subsection{Low field data}
\begin{figure*}
\begin{center}
\includegraphics[width=6.5cm,angle=-90]{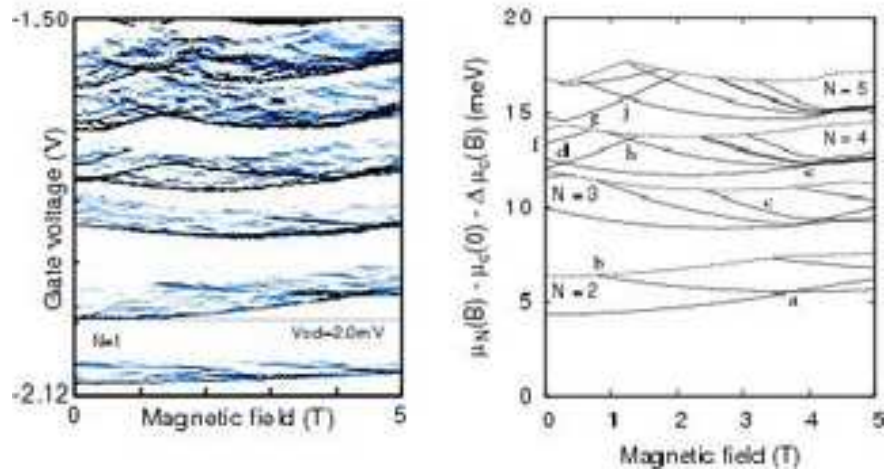}
\end{center}
\caption{Experimental and theoretical excitation spectra.
Left frame: experimental excitation spectra. The grey scale image
shows $dI_{sd}/dV_g$
in the $B, V_g$ plane for $N= 1$-6. Right frame: theoretical excitation
spectra for $N=2$-5. The dashed lines show the upper bound of the
excitation stripe, $\Delta E = 2$ meV. Roman font labels indicate the
features listed in Table~\ref{exctab}.}
\label{excfig}
\end{figure*}
\begin{table*}
\caption{Comparison of features in experimental and theoretical spectra.
Features (a), (d), (e) and (j) are ground state transitions. For these features
$(L,S)\rightarrow (L',S')$
indicates low field values of $(L,S)$ followed by high field values.
Features (b), (f), (g) and (h) are excitation energies at the points shown in
Fig.~\ref{excfig} and defined in the text.
Feature (c') is the excitation energy on line (c) at $B=3.0$ T. 
For these features $(L,S)\Rightarrow (L',S')$ indicates
excited state $(L,S)$ values followed by the ground state values.}
\begin{ruledtabular}
\begin{tabular}{lcccccr}
Feature  & $N$ & $\Delta E_{Exp}$ (meV) & $\Delta E_{The}$ (meV) &
$B_{Exp}$ (T) & $B_{The}$ (T) & $(L,S)$ values \\
\hline
(a) & 2 & 0.0 & 0.0 & 3.8 & 4.2 & $(0,0)\rightarrow (1,1)$\\
(b) & 2 & 2.0 & 2.0 & 0.75 & 1.15 & $(1,1)\Rightarrow (0,0)$\\
(c') & 3 & 1.4 & 1.22 & 3.0 & 3.0 & $(3,\frac{3}{2})\Rightarrow (1,\frac{1}{2})$\\
(d) & 4 & 0.0 & 0.0 & 0.33 & 0.3 & $(0,1)\rightarrow (2,0)$\\
(e) & 4 & 0.0 & 0.0 & 4.06 & 4.0 & $(2,0)\rightarrow (3,1)$\\
(f) & 4 & 1.0 & 1.24 & 0.0 & 0.0 & $(0,0)\Rightarrow (0,1)$\\
(g) & 4 & 2.0 & 2.0 & 0.86 & 0.70 & $(0,0)\Rightarrow (2,0)$\\
(h) & 4 & 1.5 & 1.71 & 1.57 & 1.3 & $(0,1)\Rightarrow (3,1)$\\
(j) & 5 & 0.0 & 0.0 & 1.39 & 1.25 &$(1,\frac{1}{2})\rightarrow (4,\frac{1}{2})$\\
\end{tabular}
\end{ruledtabular}
\label{exctab}
\end{table*}

The model parameters are found by fitting to the ground state addition energies
so comparison of theoretical and experimental excitation energies is a good
test of the predictive power of the model. This comparison is performed
here for the low magnetic field regime, $B\le 5$ T, to which the model has
not been applied before. A discussion of the high field regime is available
in the literature \cite{Nishi06}, see also section \ref{HiBsect}.

The experimental excitation energy data Fig.~\ref{excfig} (left panel)
consists of the derivative of the source-drain current, $dI_{sd}/dV_g$,
plotted as an intensity image in the $B, V_g$ plane. The electron number
ranges from 1 to 6. The data was collected
in a similar way to the addition energy data but during a different thermal
cycle. The source-drain
voltage $V_{sd}=2$ mV so the maximum excitation energy that can be probed  
is 2 meV. For each electron number there is a stripe that corresponds to
this 2 meV window. The width of each stripe depends on $B$ and $V_g$
because of the $B$ and $V_g$ dependence of $\alpha$. Within each strip
there are dark lines that correspond to transport through excited dot
states. The excitation energy is found by scaling the line position to the
stripe width. This gives excitation energies, $\Delta E$, to an accuracy of
about $\pm 0.1$ meV, together with field values accurate to about
$\pm 0.02$ T. In addition to the excitation lines, there is 'noise'. Some
of this is caused by emitter states which result from
density of states fluctuations in the heavily doped contacts.
To extract the $B$ dependence of the excitation lines it is necessary to
distinguish contact features from dot features. This can be
done by rejecting features that do not depend on $N$ \cite{Nishi06},
for example. However the excitation lines in Fig.~\ref{excfig} have gaps
and are of variable
intensity so the present quantitative comparison is based on features like
level crossings and crossings of excitation lines with the upper
boundaries of the excitation stripes. These features can be identified
unambiguously by comparing the topology of the experimental and theoretical
data.

The theoretical excitation energy data is shown in the right frame of
Fig.~\ref{excfig}. The figure shows $\mu_N(B) - \mu_1(0) - \Delta \mu_c(B)$
for electron numbers in the range 2 to 5. For each $N$ the energies of the
ground state and lowest four excited states are calculated and the value of
$\mu_N(B) - \mu_1(0) - \Delta \mu_c(B)$ is found for the ground state and
each of the excited
states. The excited state lines that fit into the 2 meV excitation window
are shown by the solid lines in the figure. The continuous solid line at
the bottom of each stripe is the ground state chemical potential and this
coincides with the lower boundary of each excitation window. The dashed
line at the top of each stripe shows the upper boundary, 2 meV above the
ground state line.

Qualitatively, there is very good agreement between the theory and
experiment. However some of the experimental excited state lines are not
continuous and there is some noise. The singlet-triplet transition of the
2 electron system is
clearly visible (a) and so is the excited state triplet line (b-a). For
3 electrons there are ground state transitions in the 4-5 T
interval in both the theory and experiment and an excited state line is
visible in the experimental data. This probably corresponds to line (c) in
the theory. In the case of 4 electrons there are ground state transitions
(d) and (e) in both theory and experiment. In addition, an excited state
line which has a flat maximum is present in the experimental data. This
corresponds to the crossing (h) in the theory but small symmetry
breaking effects such as disorder change the crossing into an anticrossing
and lead to a maximum in the experimental data. The second excited state
line (f-g) is also clearly visible in both the theory and experiment. For 5
electrons the ground state transition (j) has the distinctive form of a
maximum and is clearly present in both theory and experiment, and some short
excited state lines emerge from this crossing. Further, for both 4 and 5
electrons the experimental ground state line is broadened in the regions
between 4 and 5 T. This corresponds to the region where the theory predicts
that ground and excited state levels cluster together and excitation gaps
shrink. Although the agreement is generally good,
there are some discrepancies. For $N=1$ the theory predicts no excitations
within the 2 meV window at fields below 5 T but there is some structure in
the experimental data. This is thought to be caused by emitter states.
Similarly, for $N=5$ no excitations are predicted in the 2 meV window at
zero magnetic field but there is some structure in the data.

The quantitative experiment-theory comparison for features (a)-(j) is
detailed in Table~\ref{exctab}. For each feature, experimental and
theoretical values of $\Delta E$ and $B$ are given but $\Delta E = 0$ in
the case of ground and excited state transitions. The theoretical magnetic
fields are accurate to about 0.05 T, the magnetic field step used for the
calculations. The theoretical quantum numbers are also listed. In the case
of excitations ($\rightarrow$), the quantum numbers of the excited state are
shown followed by those of the ground state. In the case of ground state
transitions ($\Rightarrow$), the
quantum numbers on the low field side of the transition are followed by
those on the high field side.

In the case of the singlet-triplet transition of the 2-electron system (a),
the predicted transition field is about 0.4 T lower than in the experiment
and the point where the triplet excitation line crosses the 2 meV
excitation boundary (b) is also about 0.4 T lower. This discrepancy is
thought to be a consequence of the thermal cycling since the positions of
the singlet-triplet transition agree well in the addition energy data,
Fig~\ref{addfig2}. The excited state line of the 3-electron system (c) is
difficult to identify because it is not clear whether it extends as far as
the excitation boundary in the experimental data. However there is some
weak excited state structure in the data at around 3.5 T and this suggests
that the experimental line is the second excited state. The experimental
excitation energy is compared with the theoretical second excited state
energy at the arbitrarily chosen field of 3 T in Table~\ref{exctab}
(feature (c')). The
two values differ by only 0.18 meV and this supports the idea that the
experimental line is the second excited state. In addition, the
extrapolation of the experimental line crosses the upper boundary of the
excitation stripe at about 2.24 T compared with 2.4 T in the theory and
line (c) crosses the theoretical ground state line at 4.6 T close to an
experimental ground state transition at 4.58 T. This gives further support
to the interpretation of the experimental excited state line as the
second excited state line corresponding to theoretical line (c).

The 4-electron system is rich in features that can be identified
unambiguously and compared quantitatively with theory. Ground state
transitions at 0.33 and 4.06 T agree very well with transitions at
0.3 T (d) and 4.0 T (e) in the theory. The excitation energy of the
second excited state (f) at zero magnetic field agrees with theory to 0.24
meV while the position of the crossing of the second excited state line
with the excitation stripe boundary (g) agrees with theory to 0.16T. The
excitation energy at the excited state crossing (h) agrees to 0.21 meV and
its position agrees to 0.27 T. It is more difficult to identify features
in the 5-electron data as there is more noise. However the position of the
ground state transition (j) agrees to 0.14 T.  

\subsection{High field data}
\label{HiBsect}
Application of the present model to data in the high magnetic field regime
beyond the MDD has been discussed in
ref. \cite{Nishi06}. However the fitting procedures described here are not
detailed in that work. Here it is emphasised that identical procedures were
used in the present work and the work described in ref. \cite{Nishi06} and
that identical model parameters were used in both cases. One of the main
findings of ref. \cite{Nishi06} is the existence of intermediate spin
states in the regime beyond the MDD. The MDD is spin polarised but as the
magnetic field increases a transition to a partly polarised state occurs,
followed by a spin polarised state, then another partly polarised state and
so on. The quantum numbers found for the $N=5$ dot studied in
ref. \cite{Nishi06} are consistent with general ideas about symmetry
and electron molecular states \cite{Maksym96,Maksym00}. In the specific
case of the $N=5$ partly polarised states, calculations of the pair
correlation function indicate that there is a superposition of 4- and 5-fold
symmetry with a dominant 5-fold component. Further details are given in refs.
\cite{Maksym06,MaksymArita07}.

\section{Conclusion}
\label{ConcSection}
An accurate model of a vertical pillar quantum dot has been developed which
is able to reproduce addition energy and chemical potential data for
individual quantum dots. Experimental addition energies are reproduced to
an accuracy of around 0.15 meV and this is accurate enough to enable ground
state spin and orbital angular momentum quantum numbers to be determined
from transport data.
Although the three-dimensional device structure is accounted for, the
resulting physical model reduces to one in which electron motion is
restricted to the two lateral dimensions, the confinement is parabolic and
the interaction potential differs significantly from the bare Coulomb
potential.

The model only contains 3 adjustable parameters. Of these the
parameter that describes the $N$ dependence of the confinement energy does
not appear to be significant over the small $N$ range considered here and
it is likely that a two parameter model would give similar accuracy to the
present one. For dots with larger $N$ it would be possible to include
non-parabolic confinement. This would increase the number of model
parameters but with larger $N$ there would be more data so the additional
parameters could probably be determined reliably.

Although the model parameters are determined by fitting ground state data,
the model is able to give a good description of the low lying excited
states. The model energies agree with experiment to about 10-20\% and
the positions of features in magnetic field dependent data agree to
similar accuracy. 
One of the factors limiting the accuracy of the present analysis is
uncertainty in the value of the electrostatic leverage factor
$\alpha$. Accurate measurements of $\alpha$ would enhance the scope and
applicability of the present analysis.

The general approach described here is not limited to vertical pillar
dots. In principle, similar analysis procedures could be developed for any
type of dot, provided a good physical model is available.

\begin{acknowledgments}
ST is grateful for financial support from the Grant-in-Aid
for Scientific Research CS (No. 19104007) and B (No. 18340081), SORST-JST
and Special Coordination Funds for Promoting Science and Technology,
MEXT. PAM is grateful for hospitality at the Department of Physics,
University of Tokyo where this manuscript was completed and is also
grateful for special study leave from the University of
Leicester. We are pleased to thank H. Imamura for fruitful comments and
discussions.
\end{acknowledgments}

\appendix
\section{Numerical calculation of Green's function}
The functions $f$ and $g$ and the Wronskian $W$ (Eq.~\ref{gfform}) are
needed to find the Green's function. To find $f$ and $g$ the system is
divided into thin slices, such that $\epsilon$ and $q_0$ are constant in
each slice. In the present case this simply means that each layer of the
structure is sub-divided into thin slices but the same numerical procedure
is applicable to any form of $\epsilon(z)$. The $n$\/th slice occupies the
region between $z_{n-1}$ and $z_n$ and is characterised by a relative
permittivity $\epsilon_n$, $q$ value $q_n^2 = q^2 + q_0^2(z)$ and a
thickness $t_n = z_n - z_{n-1}$. Within
each slice $f$ and $g$ have the form $A_n\exp(-q_n z) +
B_n\exp(q_nz)$. The two terms in this expression are analogous to
evanescent waves in diffraction theory and it is convenient to use the
language of diffraction theory to describe them. Thus the reflected wave,
$R$ which decays in the positive $z$ direction has the form
$R = A_n\exp(-q_n z)$. Similarly, the transmitted wave, $T$ which
decays in the negative $z$ direction has the form
$T =B_n\exp(q_n z)$.

The amplitudes at successive slices are related via a transfer matrix,
\begin{equation}
\left(
\begin{array}{c}
R_{n+1}\\
T_{n+1}\\
\end{array}
\right)
=
\left(
\begin{array}{cc}
m_1 & m_2\\
m_3 & m_4\\
\end{array}
\right)
\left(
\begin{array}{c}
R_{n}\\
T_{n}\\
\end{array}
\right),
\label{tmatdef}
\end{equation}
where $R_n$ and $T_n$ are amplitudes just below the interface at $z_n$,
that is within the slice of relative permittivity $\epsilon_n$.
The boundary conditions on $G$ are used to find the elements of the
transfer matrix: $g$ and $\epsilon(z) dg/dz$ are continuous and the same
holds for $f$. Therefore
\begin{eqnarray}
m_1 &=& \frac{1}{2}\left(
1 + \frac{\epsilon_n q_n}{\epsilon_{n+1} q_{n+1}} \right)
\exp(-q_{n+1} t_{n+1}), \nonumber \\
m_2 &=& \frac{1}{2}\left(
1 - \frac{\epsilon_n q_n}{\epsilon_{n+1} q_{n+1}} \right)
\exp(-q_{n+1} t_{n+1}), \nonumber \\
m_3 &=& \frac{1}{2}\left(
1 - \frac{\epsilon_n q_n}{\epsilon_{n+1} q_{n+1}} \right)
\exp(q_{n+1} t_{n+1}), \nonumber \\
m_4 &=& \frac{1}{2}\left(
1 + \frac{\epsilon_n q_n}{\epsilon_{n+1} q_{n+1}} \right)
\exp(q_{n+1} t_{n+1}).
\label{tmatvals}
\end{eqnarray}

It is well known that direct application of Eq.~(\ref{tmatdef}) is
numerically unstable. Instead it is necessary to compute the ratio
$r_n = R_n / T_n$ which corresponds to the reflection coefficient at each
interface \cite{Meyer-Ehmsen89,Zhao93,Kawamura07}. The notation used here
is similar to that of ref. \cite{Kawamura07}. The reflection coefficient
$r_n$ satisfies
\begin{equation}
  r_{n+1} = \frac{m_1 r_n + m_2}{m_3 r_n + m_4},
\label{rstep}
\end{equation}
where the $m_i$ are the elements of the transfer matrix
(Eq.~(\ref{tmatvals}))
for going across the interface at $z_n$ to the interface at $z_{n+1}$.
Eq.~(\ref{rstep}) can be used to step $r_n$ provided that the denominator of
the fraction is not small. If the denominator does become small
an alternative relation can be used to step $r_n^{-1}$
\cite{Meyer-Ehmsen89} but this was not needed for the present work.

The procedure for finding $g$ is as follows. Deep inside the bottom
contact $g$ decays exponentially so the only component present is $T_n$.
Hence $r_n$ is found from Eq.~(\ref{rstep}) starting from $r_0=0$.
Once $r_n$ is known $g_n$ is found from the relations
\begin{eqnarray}
T_n &=& (m_3 r_n + m_4)^{-1} T_{n+1}, \nonumber \\
g_n &=& R_n + T_n \nonumber \\
&=& (1 + r_n) T_n,
\label{tstep}
\end{eqnarray}
which follow from Eq.~(\ref{tmatdef}) and the definitions of $R_n$ and
$T_n$. Eq.~(\ref{tstep}) is applied with the initial condition $T_N=1$,
where the topmost slice ends at $z_N$. This procedure corresponds
directly to the one used in diffraction theory.

The procedure used to find $f$ is slightly different. $f$ is required to
decay exponentially deep inside the top contact, that is, it only has a
reflected component there. This means that the inverse reflection
coefficient $T_n/R_n$ vanishes deep inside the top contact. Hence it is
convenient to re-expresss Eqs.~(\ref{rstep}) and (\ref{tstep}) in terms of
the inverse transfer matrix defined by 
\begin{equation}
\left(
\begin{array}{c}
R_{n}\\
T_{n}\\
\end{array}
\right)
=
\left(
\begin{array}{cc}
n_1 & n_2\\
n_3 & n_4\\
\end{array}
\right)
\left(
\begin{array}{c}
R_{n+1}\\
T_{n+1}\\
\end{array}
\right),
\label{itmatdef}
\end{equation}
where $R_n$ and $T_n$ are amplitudes just below the interface at $z_{n-1}$,
again within the slice of relative permittivity $\epsilon_n$.
The boundary conditions on $G$ lead to expressions for the inverse transfer
matrix elements:
\begin{eqnarray}
n_1 &=& \frac{1}{2}\left(
1 + \frac{\epsilon_{n+1} q_{n+1}}{\epsilon_n q_n} \right)
\exp(q_{n+1} t_{n+1}), \nonumber \\
n_2 &=& \frac{1}{2}\left(
1 - \frac{\epsilon_{n+1} q_{n+1}}{\epsilon_n q_n} \right)
\exp(-q_{n+1} t_{n+1}), \nonumber \\
n_3 &=& \frac{1}{2}\left(
1 - \frac{\epsilon_{n+1} q_{n+1}}{\epsilon_n q_n} \right)
\exp(q_{n+1} t_{n+1}), \nonumber \\
n_4 &=& \frac{1}{2}\left(
1 + \frac{\epsilon_{n+1} q_{n+1}}{\epsilon_n q_n} \right)
\exp(-q_{n+1} t_{n+1}).
\label{itmatvals}
\end{eqnarray}
The inverse reflection coefficient satisfies
\begin{equation}
 \tilde{r}_{n}^{-1} = \frac{n_3  + n_4 \tilde{r}_{n+1}^{-1}}
{n_1 + n_2 \tilde{r}_{n+1}^{-1}}
\label{irstep}
\end{equation}
and this relation is used to step $\tilde{r}_n^{-1}$, starting from the
initial condition $\tilde{r}_{N+1}^{-1}=0$. Here the notation $\tilde{r}$
is used to emphasise that $r$ and $\tilde{r}$ are computed with different
initial conditions so $\tilde{r}^{-1}$ is {\it not} the inverse of $r$.
Once $\tilde{r}_n^{-1}$ is known, $f_n$ is found from the relations
\begin{eqnarray}
R_{n+1} &=& (n_1 + n_2 \tilde{r}_{n+1}^{-1})^{-1} R_{n}, \nonumber \\
f_n &=& R_n + T_n \nonumber \\
&=& (1 + \tilde{r}_n^{-1}) R_n,
\label{itstep}
\end{eqnarray}
with the initial condition $R_0=1$.

The Wronskian $W$ is independent of $z$ and can be evaluated at any
convenient position. This requires $f$, $g$ and their derivatives. The
derivatives are found from
\begin{eqnarray}
\frac{df}{dz} &=& q_n (-A_n \exp(-q_n z) + B_n \exp(q_n z)) \nonumber \\
&=& q_n (T-R)
\end{eqnarray}
together with a similar expression for $dg/dz$. This leads to
\begin{equation}
W = -2q\epsilon_w\epsilon_0\frac{(1- \tilde{r}_n^{-1} r_n)}
{(1+\tilde{r}_n^{-1})(1+r_n)} f_n g_n,
\end{equation}
which is valid away from dielectric interfaces.

The relations given here have been found to be numerically stable for the
device considered in the present work. But in general it is possible
for the calculation of the Wronskian to underflow, causing overflow in the
calculation of the Green's function. However this is only likely to be a
problem for very thick systems and where underflow could result from the
exponential form of $f$ and $g$.

\section{Numerical calculation of Green's function integrals}
The Fourier components of the effective interaction are found from
Eq.~(\ref{u2dqdef}). The form factor is
\begin{equation}
F(q) = 2q \epsilon_w \epsilon_0 \int \chi^2(z) \chi^2(z') G({q}, z, z') dz dz'.
\end{equation}
Except for a factor of $-1/W$, the integral in this
equation has the form
\begin{eqnarray}
&&\int_{-\infty}^{\infty} \chi^2(z) f(z)
\int_{-\infty}^{z} \chi^2(z') g(z') dz' dz \nonumber \\
&+&
\int_{-\infty}^{\infty} \chi^2(z) g(z)
\int_{z}^{\infty} \chi^2(z') f(z') dz' dz \nonumber \\
&=&
2 \int_{-\infty}^{\infty} \chi^2(z) f(z)
\int_{-\infty}^{z} \chi^2(z') g(z') dz'dz.
\label{doubleint}
\end{eqnarray}
To evaluate this integral numerically it is convenient to define
\begin{equation}
I(z) = f(z) \int_{-\infty}^{z} \chi^2(z') g(z') dz'.
\end{equation}
Then $I(z)$ is evaluated recursively from
\begin{eqnarray}
&&I(z+\Delta z) = \frac{f(z+\Delta z)}{f(z)} I(z) \nonumber \\
&+& f(z+\Delta z)
\int_{z}^{z + \Delta z} \chi^2(z') g(z') dz'
\end{eqnarray}
and the required integral is found from
\begin{equation}
\int_{-\infty}^{\infty} \chi^2(z) I(z) dz.  
\end{equation}
The advantage of this approach is that only a few operations are needed to
update $I(z)$. As a result the computer time scaling is linear with the
number of integration steps instead of quadratic as would be the case if
the integral in Eq.~(\ref{doubleint}) was evaluated directly.

The approach is simple to implement on a uniform grid. However a uniform
grid cannot be used in the present case because the system contains
interfaces where the relative permittivity changes abruptly and it is
necessary to ensure that the grid points coincide with these
interfaces. This is done by adjusting the step length within each region
of constant relative permittivity. As a result the step length is
constant within each region but the step length varies from region to
region. To apply the approach to the resulting non-uniform grid the
following generalisation of Simpson's rule is used:
\begin{equation}
\int^{z_0 + \Delta_2}_{z_0 - \Delta_1} u(z) dz \sim
a u(z_0-\Delta_1) + b u(z_0) + c u(z_0+\Delta_2),
\end{equation}
where $a,b,c$ are chosen so that the integration rule is exact for
$u(z)= 1, z, z^2$ and $\Delta_1$ and $\Delta_2$ are step lengths.
The values of $a,b,c$ are
\begin{eqnarray}
a &=& \frac{2\Delta_1^2+\Delta_1\Delta_2-\Delta_2^2}{6\Delta_1}, \nonumber \\
b &=& \frac{3(\Delta_1 + \Delta_2)\Delta_1\Delta_2 +\Delta_1^3 + \Delta_2^3}
{6\Delta_1\Delta_2}, \nonumber \\
c &=& \frac{2\Delta_2^2+\Delta_1\Delta_2-\Delta_1^2}{6\Delta_2}.
\end{eqnarray}

\end{document}